\documentstyle[12pt]{article}

\catcode`@=11
\def\secteqno{\@addtoreset{equation}{section}%
\def\theequation{\thesection.\arabic{equation}}}
\catcode`@=12
\secteqno
\newcommand{\be}{\begin{equation}}
\newcommand{\ee}{\end{equation}}
\newcommand{\bea}{\begin{eqnarray}}
\newcommand{\eea}{\end{eqnarray}}
\newcommand{\bref}[1]{(\ref{#1})}

\catcode`@=11

\begin{document}
\vfill
\today
\vbox{
\hfill 
}\null

\vskip 10mm
\begin{center}
{\Large\bf Mass Matrix Model Broken From  A4 To 2 $\leftrightarrow$ 3 Symmetry
\\
}\par
\vskip 10mm
{Takeshi FUKUYAMA}\par
\medskip
Department of Physics, Ritsumeikan University,\\ 
Kusatsu, Shiga, 525-8577 Japan

\medskip
\vskip 10mm
\end{center}
\vskip 10mm
\begin{abstract}
2$\leftrightarrow$3 symmetry is realized by the breaking from alterating group of degree 4 ($A4$) symmetry.
$A4$ explains why the generation number is three.
However the mass matrices are realized in the form of the breaking to $2\leftrightarrow 3$ symmetry $\times Z_3$, which leads us to $2\leftrightarrow 3$ symmetric mass matrix with vanishing (1,1) component. 
Thus the $3\times 3$ mass matrix model with $2\leftrightarrow 3$ symmetry and vanishing (1,1) component has the group theoretical background as the symmetry in GUT model.

                  
\end{abstract}

\vskip 2cm

\section{Introduction}
There have been many discussions on GUTs in these decades.
These study the interrelations between quarks and leptons mass matrices.
For instance, the renormalizable minimal SO(10) GUT predicts the mass relations \cite{Babu} \cite{Matsuda}
\begin{eqnarray}
  M_u &=& c_{10} M_{10} + c_{126} M_{126}   \nonumber \\
  M_d &=&     M_{10} +     M_{126}   \nonumber \\
  M_D &=& c_{10} M_{10} -3 c_{126} M_{126}   \nonumber \\
  M_e &=&     M_{10} -3     M_{126}   \nonumber \\
  M_L &=& c_L M_{126} \nonumber \\  
  M_R &=& c_R M_{126}  \; , 
 \label{massmatrix}
\end{eqnarray} 
Here $M_u,~ M_d,~M_D,~M_e,~M_L,~M_R$ are $3\times 3$ up type quark, down  type quark, Dirac neutrino, charged lepton, left-handed Majorana, and right-handed Majora neutrino mass matriices, respectively. They are composed of only two kinds of mass matrices $M_{10}$ and $M_{126}$. Thus SO(10) group properties predict the combination of $M_{10}$ and $M_{126}$ between quark lepton mass matrices but do not predict any structure of $M_{10}$ and $M_{126}$ themselves.
Interrelations between different families (so called family symmetry) become
 the recent topics.
One of the most important problems of family symmetries may be why we have three families.
This may be partly solved by $A4$ group since it leads us to triplet Higgs \cite{Altarelli}. A4 is the four degreed symmetry group with even permutation whose elements we denote as $(a_1,a_2,a_3,a_4)$. $A4$ is generated by the $S$ and $T$ and their products, which satisfy
\be
S^2=T^3=(ST)^3=1
\ee
The three-dimensional unitary representation, in a basis
where the element $S$ is diagonal, is built up from:
\begin{equation}\label{tre}
S=\left(\matrix{
1&0&0\cr
0&-1&0\cr
0&0&-1}\right),~
T=\left(\matrix{
0&1&0\cr
0&0&1\cr
1&0&0}
\right).
\ee
Let us practice the transformation, for instance, $ST$ to $V\equiv (a_1,a_2,a_3)^T$
\be
TV=\left(
\begin{array}{l}
a_2\\
a_3\\
a_1\\
\end{array}
\right)
\ee
\be
STV=\left(
\begin{array}{l}
a_2\\
-a_3\\
-a_1\\
\end{array}
\right)
\ee
The rule of the game for reading the permutation group of four degree from three dimensional vector is to make plus element change to $a_4$ and do minus signs interchange, and $ST$ corresponds to $(a_4,a_1,a_3,a_2)$.
Thus $S$ means the $2\leftrightarrow 3$ symmetry and $T$ does cyclic permutation or equivalently $Z_3$.

Mathematically this is the elementary example of Sylow's theorem \cite{Huppert}:
The order of $A4$ is $12=2^2\times 3$, and it is the product of normal subgroup $V_4$, composed of
\be
(1,2)(3,4),~(1,3)(2,4),~(1,4)(2,3),~1,
\ee
and $Z_3$.
Thus $A4\supset (2\leftrightarrow 3) \mbox{symmetry} \times Z_3$.
This fact is very important for the model buiding as will been shown soon.

Unfortunately $A4$ is too restrictive for the universal structures of quark and lepton mass matrices.
$2\leftrightarrow 3$ symmetry was first proposed by us as the neutrino mass matrix model \cite{F-N} and soon be extended to quark sectors by incorpolating CP phases \cite{Koide}.
\subsection{$Z_3$ symmetry}
Let us start from $Z_3$ symmetry.
We assign $Z_3$ charge of each generation of fermions as
\be
\begin{array}{c}
\psi_{1L} \rightarrow \psi_{1L} , \\
\psi_{2L} \rightarrow \omega \psi_{2L} , \\
\psi_{3L} \rightarrow \omega \psi_{3L}, \\
\end{array}
\ee
where $\omega^3=+1$.
Then, the bilinear terms $\overline{q}_{Li} u_{Rj}$, 
$\overline{q}_{Li} d_{Rj}$, $\overline{\ell}_{Li} \nu_{Rj}$, 
$\overline{\ell}_{Li} e_{Rj}$ and $\overline{\nu}_{Ri}^c \nu_{Rj}$
[$\nu_R^c =(\nu_R)^c =C \overline{\nu_R}^T$ and
$\overline{\nu}_R^c =\overline{(\nu_R^c)}$] 
are transformed as follows:
\begin{equation}
\left( 
\begin{array}{ccc}
1 & \omega^2 & \omega^2 \\
\omega^2 & \omega & \omega \\
\omega^2 & \omega & \omega \\
\end{array} \right) \ ,
\end{equation} 
where
\begin{equation}
q_{L} =\left(
\begin{array}{c}
u_L \\
d_L
\end{array} \right) \ , \ \ \ 
\ell_{L} =\left(
\begin{array}{c}
\nu_L \\
e^-_L
\end{array} \right) \ .
\end{equation}
Therefore, if we assume two SU(2) doublet Higgs scalars $H_1$ and $H_2$ (as we will soon argue, this is the case of realisitic models), 
 which are transformed as
\be
H_1 \rightarrow \omega H_1 , \ \ \ H_2 \rightarrow \omega^2 H_2 , 
\ee
the Yukawa interactions are given  as follows
\begin{eqnarray}
H_{{\rm int}} &=&
 \sum_{A=1,2}\left( Y^{u}_{(A)ij}\overline{q}_{Li}\widetilde{H}_{A} u_{Rj}
+ Y^{d}_{(A)ij}\overline{q}_{Li}{H}_A d_{Rj} \right)\nonumber \\
&+&  \sum_{A=1,2}\left( Y^{\nu}_{(A)ij}\overline{\ell}_{Li}\widetilde{H}_{A} 
\nu_{Rj}
+ Y^{e}_{(A)ij}\overline{\ell}_{Li}{H}_A e_{Rj} \right) \\
&+& \left( Y^{R}_{(1)ij}\overline{\nu}_{Ri}^c \widetilde{\Phi}^0 \nu_{Rj} 
   + Y^{R}_{(2)ij}\overline{\nu}_{Ri}^c {\Phi}^0 \nu_{Rj} \right)
+ {\rm h.c.} \ ,\nonumber
\end{eqnarray}
where
\begin{equation}
H_A =\left(
\begin{array}{c}
H^+_A \\
H^0_A
\end{array} \right) \ , \ \ \ 
\widetilde{H}_A =\left(
\begin{array}{c}
\overline{H}^0_A \\
-H^-_A
\end{array} \right) \ , 
\label{tilde}
\end{equation}

Therefore,
\be
Y^u_{(1)},\ Y^d_{(2)},\ Y^\nu_{(1)},\ Y^e_{(2)},\ Y^R_{(2)} =
\left(
\begin{array}{ccc}
0 & 0 & 0 \\
0 & \ast & \ast \\
0 & \ast & \ast \\
\end{array}
\right)
\ , \ \ \ \ 
Y^u_{(2)},\ Y^d_{(1)},\ Y^\nu_{(2)},\ Y^e_{(1)},\ Y^R_{(1)} =
\left(
\begin{array}{ccc}
0 & \ast & \ast \\ 
\ast & 0 & 0 \\ 
\ast & 0 & 0 \\
\end{array}
\right).
\label{ast}
\ee
In \bref{ast}, the symbol $*$ denotes  non-zero quantities. 
Here, in order to give  heavy Majorana masses of the right-handed neutrinos
$\nu_R$, we have assumed an SU(2) singlet Higgs scalar $\Phi^0$, which is 
transformed  as $H_1$.
Mass matrices are sum of $Y_{(1)}$ and $Y_{(2)}$ and their (1,1) element must be vanished:
\be
\left(
\begin{array}{ccc}
0 & \ast & \ast \\ 
\ast & \ast & \ast \\ 
\ast & \ast & \ast \\
\end{array}
\right).
\label{ast2}
\ee
For the minimal SO(10) GUT two sets of doublets come from
$({\bf 1,2,2})\subset {\bf 10}$ and $({\bf \overline{15},2,2})\subset \overline{{\bf 126}}$ under $SU(4)_c\times SU(2)_L\times SU(2)_R$, and Yukawa coupling is
\begin{eqnarray}
W_Y &=& 
\overline{u}_i  \left(
 Y_{10}^{ij}  H^u_{10} + Y_{126}^{ij}  H^u_{126}     
 \right) q_j 
+
 \overline{d}_i  \left(
 Y_{10}^{ij}  H^d_{10} + Y_{126}^{ij}  H^d_{126}     
\right) q_j  \nonumber \\ 
&+&
 \overline{\nu}_i  \left(
 Y_{10}^{ij}  H^u_{10} - 3 Y_{126}^{ij} H^u_{126}   
\right) \ell_j 
+
\overline{e}_i  \left(
 Y_{10}^{ij}  H^d_{10}  - 3 Y_{126}^{ij} H^d_{126}  
\right) \ell_j   \nonumber \\
&+&
\overline{\nu}_i  
 \left( Y_{126}^{ij} \; v_R \right) 
\overline{\nu}_j \;  , 
\label{Yukawa2}
\end{eqnarray} 
Thus in this case we have not two Higgs doublets but two sets of Higgs doublets, 

($H_1=H_{10}^u,~H_2=H_{126}^u$) and ($\widetilde{H_1}=H_{10}^d,~\widetilde{H_2}=H_{126}^d$).
However the transformation properties are same and the above discussions remains valid. $v_R$ in \bref{Yukawa2} is the vev of $({\bf 15,1,3})\subset {\bf 126}$, a singlet under Standard gauge group.

The remaining $2\leftrightarrow 3$ symmetry gives the constraint on $\ast$ in \bref{ast2}.

\subsection{$2\leftrightarrow 3$ symmetry}
 $2\leftrightarrow 3$ symmetry in addition to $Z_3$ gives the constraint to mass matrix \cite{Koide}
\indent
$\widehat{M_f}$  as
\begin{equation}
\widehat{M_f}=\left(
\begin{array}{lll}
\ 0 & \ A_f & \ A_f \\
\ A^\prime_f & \ B_f & \ C_f \\
\ A^\prime_f & \ C_f & \ B_f \\
\end{array}
\right), \ \ 
\end{equation}
where $A_f$, $A^\prime_f$, $B_f$, and $C_f$ are real parameters. 

This matrix is block diagonalized by the maximal mixing $O'$ between
2'nd and 3'rd generation
\be
O'=\left(
\begin{array}{lll}
1 & 0 & 0\\
0 & \sqrt{\frac{1}{2}} & -\sqrt{\frac{1}{2}}\\
0 & \sqrt{\frac{1}{2}} & \sqrt{\frac{1}{2}}\\
\end{array}
\right)
\ee
as
\be
O'^T\widehat{M_f}O'=\left(
\begin{array}{lll}
0 & \sqrt{2}A & 0\\
\sqrt{2}A & B+C & 0\\
0 & 0& B-C\\
\end{array}
\right)
\ee
This is diagonalized  by two orthogonal matrices $O_{f1}$ and $O_{f2}$ as 
\begin{equation}
O_{f1}^T \widehat{M_f} O_{f2} = \mbox{diag}(m_{f1} , m_{f2} , m_{f3}),
\end{equation}
where $m_{f1}$, $m_{f2}$, and $m_{f3}$ are eigenvalues of $M_f$. The orthogonal matrices $O_{f1}$ and $O_{f2}$ are given by 
\begin{equation}
O_{f1} = O'\left(
\begin{array}{ccc}
cos\varphi_{f1} & -sin\varphi_{f1} &  0 \\
sin\varphi_{f1} & cos\varphi_{f1} &  0 \\
0 & 0 & 1
\end{array}
\right).\label{O_f_1}
\end{equation}
\begin{equation}
O_{f2} = O' \left(
\begin{array}{ccc}
cos\varphi_{f2} & -sin\varphi_{f2} &  0 \\
sin\varphi_{f2} & cos\varphi_{f2} &  0 \\
0 & 0 & 1
\end{array}
\right),\label{O_f_2}
\end{equation}
If $\varphi_{f1}=\varphi_{f2}=-\pi/6$, mixing gives tribimaximal. 
\begin{equation}
O = \left(
\begin{array}{ccc}
 \frac{2}{\sqrt{6}} &  \frac{1}{\sqrt{3}} &  0 \\
- \frac{1}{\sqrt{6}}  & \frac{1}{\sqrt{3}}  &  - \frac{1}{\sqrt{2}}  \\
- \frac{1}{\sqrt{6}} & \frac{1}{\sqrt{3}}  & \frac{1}{\sqrt{2}} 
\end{array}
\right).\label{tribimaximal}
\end{equation}
That is, the maximal subgroup $2\leftrightarrow 3$ symmetry is the symmetric matrix $A^\prime_f =A_f$, and therefore $O_{f1}=O_{f2}$. 
That is, $2\leftrightarrow 3$ symmetry proposed in \cite{F-N} is formulated in the subgroup of $A4$ model.
As is well known, $A4$ symmetry is considered as the model of lepton sector and
has not found the successful generalization to quark sector.
If we incorporate the CP phases which is out of $A4$ arguments, we can give the consistent predictions of quark mass matrices as well as of lepton sectors \cite{Koide}
\section{Acknowledgement}

The author is grateful to S.F. King, H.Ishii, K.Matsuda, and H. Nishiura for useful comments.

\end{document}